\documentclass[aps,prd,preprint,superscriptaddress,amsmath,amssymb,showpacs,nofootinbib]{revtex4-1}
\usepackage{dcolumn}
\usepackage{graphicx}
\usepackage{float}
\usepackage{physics}
\usepackage[colorlinks=true,allcolors=blue]{hyperref}
\usepackage{mathrsfs}
\usepackage{inputenc}
\usepackage{appendix}
\usepackage[T1]{fontenc}
\usepackage{babel}
\usepackage{geometry}
\begin{document}

\title{De Haas - van Alphen Effect under Rotation}
\author{Shu-Yun Yang}
\thanks{yangsy@mails.ccnu.edu.cn}
\affiliation{ Institute of Particle Physics and Key Laboratory of Quark and Lepton Physics (MOS), Central China Normal University, Wuhan 430079, China}

\author{Ren-Da Dong}
\thanks{}

\author{De-Fu Hou}
\thanks{Co-corresponding author: houdf@mail.ccnu.edu.cn}
\affiliation{ Institute of Particle Physics and Key Laboratory of Quark and Lepton Physics (MOS), Central China Normal University, Wuhan 430079, China}

\author{Hai-Cang Ren}
\thanks{Co-corresponding author: renhc@mail.ccnu.edu.cn}
\affiliation{ Physics Department, The Rockefeller University, 1230 York Avenue, New York, NY 10021-6399}
\affiliation{ Institute of Particle Physics and Key Laboratory of Quark and Lepton Physics (MOS), Central China Normal University, Wuhan 430079, China}

\begin{abstract}
 We explored the interplay between magnetic field and rotation in the de Hass - van Alphen oscillation. The effect is found to be reduced because of the re-weighting of different states within the same Landau level by rotation energy. The implications of our results
 on high energy physics and condensed matter physics are speculated.

\end{abstract}

\maketitle
\section{Introduction}

The experimental activities for recent years regarding the polarization~\cite{STAR:2017ckg,STAR:2007ccu,STAR:2018gyt,STAR:2019erd,STAR:2020igu,STAR:2020xbm,STAR:2021beb} and chiral magnetic effects~\cite{STAR:2021pwb,STAR:2020gky,STAR:2021mii} in off-central relativistic heavy ion collisions promoted theoretical research interests in a rotating thermodynamic system in a magnetic field~\cite{Becattini:2021lfq,Chen:2015hfc,Fukushima:2020ncb,PhysRevLett.117.152002,Liu:2017zhl,Mottola:2019nui}. The same physical conditions are also present in a neutron star~\cite{Felipe:2007vb,Watanabe:2022cuv,Jerome:2022emr,Chatterjee:2021wsr}. One of the inteplay between the magnetism and rotation, the Barnett effect (or Einstein-de Haas effect)~\cite{einstein1915verh,barnett1935gyromagnetic,Bhadury:2022ulr} has been considered in hydrodynamic modeling of the collisions. In this work, we examine another interplay between magnetism and rotation, i.e. the de Haas - van Alphen effect~\cite{ONUKI20014964,Jiang:2006xe} in a strongly degenerate rotating Fermi gas. Though purely theoretical at present stage, the implications are expected to shed light on the magnetic properties of the quark matter core, if exists, in a neutron star and/or the QGP droplet of generated in the RHIC STAR fixed target experiment, where the quark density is towards the strong degeneracy. The conclusion may also be tested directly in condensaed matter physics.

De Haas-van Alphen effect is the consequence of charged fermions filling discrete but highly degenerate Landau levels~\cite{Cangemi:1996tp,Zhang:2020ben} in a magnetic field. In the absence of rotation, all degenerate Landau levels are equally populated at thermal equilibrium and the disceteness of different Laudau level is refelected in the thermodynamic limit as the oscillatory terms with respect to the chemical potential and the magnetic field in the thermodynamic potential, magnetization and magnetic susceptibility as well as some transport coefficients. When the system is in rotation, the thermodynamic equilibrium is established under a nonzero macroscopc angular momentum. The equal distribution of different angular momentum states within a Laudau level is offset by the nonzero angular velocity with higher angular momenta more favored than lower ones, which amounts to lifts the degeneracy of the Landau level. The dHvA oscillation is thereby expected to be reduced by the rotation. Consider a cylindrical volume of radius $R$ with a constant magnetic field parallel to its axis, the states of each fermion is characterized by the $z$-component of the momentum $q$, the $z$-component of the angular momentum $M$, and the radial quantum number of the wave function, $n(\ge 0)$. A Laudau levels corresponds $M>0$, the cyclotron motion in classical picture, and all $M>0$ are degenerate up to $M\sim eBR^2$, when the cyclotron orbit reaches the boundary. While the energy of a Landau level depends only on $q$ and $n$. The nonzero angular velocity $\omega$ weight different M differently through the Boltzmann factor $e^{M\omega/T}$ in the ensemble of a macroscopic angular momentum. On the other hand, the requirement of subluminal linear speed on the boundary limits the radius of the cylinder $R<1/\omega$ and the thermodynamic limit $R\to\infty$ is unrealistic and the degeneracy of the Landau levels becomes finite. We shall take the thermodynamic approximation by retaining the leading term in power in $1/R$ in the thermodynamic potential, keeping in mind $\omega R=O(1)$\footnote{In this case the kinetic energy of rotation grows with the volume, like other extensive thermodynamic quantities.}, and a sharp cutoff in the summation over angular momentum states within a Landau level is introduced to tak care of the finite size effect of the spectrum. Consequently, the implication of the rotation in the dHvA oscillation dependes on the size of the size of the system and the angular velocity. As we shall see, the dHvA is completely suppressed for typical parameters appropriate in a neutron star but may lead to observaservable effect for a cold and dense QGP fire ball created in future RHIC project. For a strongly degenerate non-relativistic electron gas, the reduction of the dHvA may be detectable in a rotating metallic sample.

This paper is organized as follows. In section II, the dHvA term of an rotating ultra-relativistic quark gas is calculated and its implications is discussed. The same effect for a non-relativistic electron is examined in section III. Section IV concludes the paper.


\section{Ultra Relativistic Fermi Gas}
\subsection{Solution of Dirac Equation in Cylindrical Cooredinate}
For a massless fermion of electric charge $e$ in a constant magnetic field $\vec B=B\hat z$ reads, the Hamiltonian in chiral representation reads
\begin{eqnarray}
	H=-i\vec{\alpha}\cdot(\vec{\nabla}-ieA)=\left(\begin{array}{cc}
		-i\vec{\sigma}\cdot(\vec{\nabla}-ieA) & 0\\
		0 & i\vec{\sigma}\cdot(\vec{\nabla}-ieA)
	\end{array}\right)
\end{eqnarray}
where the vector potential
\begin{equation}
\vec{A}=\frac{1}{2}\vec{B}\times\vec{r}
\end{equation}
We adapt the circular gauge instead of Landau gauge for the convenience of investigating a rotating Fermi gas. As the fermions of opposite chiralities have identical spectrum, we shall focus one of them in what follows with the Hamiltonian
\begin{equation}
H=-i\vec{\sigma}\cdot(\vec{\nabla}-ieA)
\label{hamiltonian}
\end{equation}
and the eigenvalue equation $H\chi=E\chi$.
For the ansatz of the two-component wave function $\chi$ in cylindrical coordinates, i.e.
\begin{eqnarray}
\chi(\vec r)=\left(\begin{array}{c}
f(\rho)e^{i\left(M-\frac{1}{2}\right)\phi}\\
g(\rho)e^{i\left(M+\frac{1}{2}\right)\phi}
\end{array}\right)e^{iqz}
\label{solution}\end{eqnarray}
we have the equations for the radial functions $f(\rho)$ and $g(\rho)$
\begin{eqnarray}
\begin{cases}
q\ensuremath{f(\rho)}-i\left(\frac{d}{d\rho}+\frac{M+\frac{1}{2}}{\rho}-\frac{1}{2}eB\rho\right)\ensuremath{g(\rho)}=E\ensuremath{f(\rho)} & \\
-i\left(\frac{d}{d\rho}-\frac{M-\frac{1}{2}}{\rho}+\frac{1}{2}eB\rho\right)\ensuremath{f(\rho)}-qg(\rho)=Eg(\rho) &
\end{cases}
\label{weyl}
\end{eqnarray}
where, $q$ and $M$ are the eigenvalue of the momentum and total angular momentum in the direction of the magnetic field with $M=\pm 1/2, \pm 3/2, ...$. The equation (\ref{weyl}) can be solved in terms of the generaliized Laguerre polynomial $L_n^\mu(z)$ and we end up with the normalized wave function~\cite{Fang:2021mou},
\begin{eqnarray}
\chi_{nMqs}(\vec r)=\frac{1}{2\pi}\sqrt{\frac{n!}{(n+m)!}}e^{-\frac{\zeta}{2}}\left(\begin{array}{c}
\sqrt{\frac{eB(E+q)}{2E}}\zeta^{\frac{m}{2}}L_n^m(\zeta)e^{im\phi}\\
\frac{iseB}{\sqrt{E(E-q)}}\zeta^{m+1}L_{n-1}^{m+1}(\zeta)e^{i(m+1)\phi}
\end{array}\right)e^{iqz}
\label{solution_1} \end{eqnarray}
for $M > 0$, and
\begin{eqnarray}
\chi_{nMqs}(\vec r)=\frac{1}{2\pi}\sqrt{\frac{n!}{(n+|m|)!}}e^{-\frac{\zeta}{2}}\left(\begin{array}{c}
\sqrt{\frac{eB(E+q)}{2E}}\zeta^{\frac{|m|}{2}}L_n^{|m|}(\zeta)e^{im\phi}\\
-\frac{iseB(n+|m|)}{\sqrt{E(E-q)}}\zeta^\frac{(|m|-1)}{2}L_n^{|m|-1}(\zeta)e^{i(m+1)\phi}
\end{array}\right)e^{iqz}
\label{solution_2} \end{eqnarray}
for $M < 0$, where $\zeta\equiv\frac{1}{2}eB\rho^2$, $m\equiv M-1/2$, $n=0,1,2,...$ and $s=\pm$. The corresponding eigenvalue of energy is $E=sE_{nMq}$ with
\begin{eqnarray}
E_{nMq}=\begin{cases}
\sqrt{2neB+q^2} & \hbox{for $M>0$}\\
\sqrt{2(n+|m|)eB+q^2} & \hbox{for $M<0$}
\end{cases}  \label{landau}
\end{eqnarray}
Care must be exercised for the case $n=0$ of the solution (\ref{solution_1}) because of the nonexistence of
$L_{-1}^{m+1}$ and the sigularity at $E=-q$. For $E=\pm q$, eq.(\ref{weyl}) becomes
\begin{eqnarray}
\begin{cases}
	\left(\frac{d}{d\rho}+\frac{m+1}{\rho}-\frac{1}{2}eB\rho\right)\ensuremath{g(\rho)}=i(\pm q-q)\ensuremath{f(\rho)}\\
		\left(\frac{d}{d\rho}-\frac{m}{\rho}+\frac{1}{2}eB\rho\right)\ensuremath{f(\rho)}=i(\pm q+q)g(\rho)
\end{cases}
\end{eqnarray}
A normalizable solution exists only if $E=q$ and reads
\begin{equation}
\chi_{0Mqs}(\vec r)=\frac{2^{m+1}}{\sqrt{\pi}}(eB)^{\frac{m+1}{2}}\rho^me^{-\frac{1}{4}eB\rho^{2}+im\phi+iqz}\left(\begin{array}{c}
1\\
0\end{array}\right)
\end{equation}
with $s={\rm sign}(q)$, which implies up(down) mover for positive(negative) energy solution.
The wave function (\ref{solution_2}) corresponds to the classical motion along the cyclotron orbit and the spectrum (\ref{landau}) constitues the entire set of Landau levels and is responsible to magnetic properties including de Haas - van Alphen effect to be discussed below in thermodynamic approximation. The wave function (\ref{solution_2}) and the spectrum (\ref{landau}) is specific to the cylindrical coordinates and is subleading in the thermodynamic approximation as we shall see below.

\subsection{Thermodynamic Pressure}

The Hamiltonian of massless fermion field in a magnetic filed is given by
\begin{eqnarray}
\mathcal{H} = \int d^{3}\vec{r}\psi^{\dagger}H\psi
\end{eqnarray}
where $H$ the single particle Hamiltonian (\ref{hamiltonian}) and the field operator
\begin{eqnarray}
\psi(\vec r)=\sum_{nMq}\eta_{nM}(q)(a_{nMq}\chi_{nMq+}(\vec r)+b_{nM-q}^\dagger\chi_{nMq-}(\vec r))
\end{eqnarray}
where
\begin{eqnarray}
\eta_{nM}(q)=\begin{cases}
\theta(q) & \hbox{for $M>0$ and $n=0$}\\
1 & \hbox{otherwise}
\end{cases}
\end{eqnarray}
We have
\begin{eqnarray}
\mathcal{H}=\sum_{n,M,q}\eta_{nM}(q)E_{nMq}(a_{nMq}^\dagger a_{nMq}
+b_{nMq}^\dagger b_{nMq})
\end{eqnarray}
Correspondingly, the fermion number operator
\begin{eqnarray}\begin{aligned}
Q=&\int d^3\vec r\psi^\dagger\psi \\
=&\sum_{n,M,q}\eta_{nM}(q)(a_{nMq}^\dagger a_{nMq}-b_{nMq}^\dagger b_{nMq})
\end{aligned}\end{eqnarray}
and the angular momemtum projection operator
\begin{eqnarray}\begin{aligned}
J_z=&\int d^3\vec r\psi^\dagger\left(-i\frac{\partial}{\partial\phi}+\frac{1}{2}\sigma_z\right)\psi \\
=&\sum_{n,M,q}\eta_{nM}(q)M(a_{nMq}^\dagger a_{nMq}-b_{nMq}^\dagger b_{nMq})
\end{aligned}\end{eqnarray}
Consequently, the thermodynamic pressure at temperature $T$ and chemical potential $\mu$ of a system rotating about $z$-axis with an angular velocity $\omega$ is
\begin{eqnarray}
	\begin{aligned}
P=&\frac{T}{\Omega}\sum_{n=0,M>0,q>0}[\ln(1+e^{-\beta(|q|-M\omega-\mu)})+\ln(1+e^{-\beta(|q|+M\omega+\mu)})] \nonumber \\
&+\frac{T}{\Omega}\sum_{n=0,M>0,q}[\ln(1+e^{-\beta(\sqrt{q^{2}+2neB}-M\omega-\mu)})+\ln(1+e^{-\beta(\sqrt{q^{2}+2neB}+M\omega+\mu)})] \nonumber \\
&+\frac{T}{\Omega}\sum_{n\neq0,M>0,q}[\ln(1+e^{-\beta(\sqrt{q^{2}+2(n+M+\frac{1}{2})eB}+M\omega-\mu)})\\
&+\ln(1+e^{-\beta(\sqrt{q^{2}+2(n+M+\frac{1}{2})eB}-M\omega+\mu)})] \nonumber
\end{aligned}
\label{pressure}
\end{eqnarray}
where we have switched the sign of $M$ of the lower branch of the spectrum (\ref{landau}) for clarity.
For a cylinder of radius $R$ and length $L$, $\Omega=\pi R^2L$,
\begin{equation}
\sum_{n,M,q}(...)=\frac{1}{\pi R^2}\int_{-\infty}^\infty\frac{dq}{2\pi}\sum_{n,M}(...)
\end{equation}
To avoid superluminal linear speed on the boundary, we require $v\equiv\omega R<1$. So the true thermodynamic limit $R\to\infty$ is not attainable but we may still take the thermodynamic approximation for sufficiently large $R$ by sorting the terms according to its power keeping in mind that $\omega R=O(1)$. For a finite $R$ summation over $M$ is limited. If follows from eqs. (\ref{solution_1}) and (\ref{solution_2}) that the square of the wave function for large $M$ and finite $n$ is peaked at the maximum of $\rho^{2|m|}\exp\left(-\frac{1}{2}eB\rho^2\right)$, which gives rise to $\rho^2=2|m|/(eB)$. When this $\rho$ becomes comparable with $R$ the finite size effect will distore the spectrum (\ref{landau}). Therefore, we introduce a cutoff for the summation over $M$, i.e.
\begin{equation}
M\leq M_c=[\frac{1}{2}eBR^2]>>1
\label{cutoff}
\end{equation}
with $[...]$ tuncate the argument inside to its integer part. As will be shown below, this cutoff produces the dHvA effect obtained from the Landau gauge in the absence of rotation. Without solving the boundary value problem of the edge states, we assume the uncertainty $\delta M_c=O(1)$ of the cutoff.

Assuming strong degeneracy, $\mu>>T$, the antiparticle contributions may be ignored \footnote{To be cautious, let us examine whether the combination ${\mathcal E}\equiv\sqrt{q^{2}+2(n+M+\frac{1}{2})eB}-M\omega$ in the last term of (\ref{pressure}) can become negative and compete with $\mu$ for large $M$. For the maximum $M(=M_c)$, ${\mathcal E}>\sqrt{2M_ceB}-M_c\omega\simeq eBR(1-v/2)>0$. The approximation of dropping the antiparticle contribution appears safe.} and we end up with
\begin{eqnarray}
	\begin{aligned}
P=&\frac{T}{\pi R^2}\int_0^\infty\frac{dq}{4\pi}\sum_{M>0}\ln(1+e^{-\beta(|q|-M\omega-\mu)})+\frac{T}{\pi R^2}\int_{-\infty}^\infty\frac{dq}{2\pi}\sum_{n>0,M>0} \ln(1+e^{-\beta(\sqrt{q^2+2neB}-M\omega-\mu)}) \\
&+\frac{T}{\pi R^2}\int_{-\infty}^\infty\frac{dq}{2\pi}\sum_{n,M>0}\ln(1+e^{-\beta(\sqrt{q^{2}+2(n+M+\frac{1}{2})eB}+M\omega-\mu)})
\end{aligned} \label{p}
\end{eqnarray}
where the contribution of the lowest Landau level has been isolated from higher Landau levels because different integration domain of $q$. The summation over $M$ in the third term of (\ref{p}) converges in the limit $M_c\to\infty$ and thereby does not contribute to the thermadynamic limit and we are left with the Landau level terms only, i.e.
\begin{eqnarray}
	\begin{aligned}
P=&\frac{T}{\pi R^2}\int_0^\infty\frac{dq}{4\pi}\sum_{M>0}\ln(1+e^{-\beta(|q|-M\omega-\mu)})+\frac{T}{\pi R^2}\int_{-\infty}^\infty\frac{dq}{2\pi}\sum_{n>0,M>0}\ln(1+e^{-\beta(\sqrt{q^2+2neB}-M\omega-\mu)})   \\
\equiv&\frac{1}{\pi R^2}P_M
\end{aligned}
\end{eqnarray}
where
\begin{equation}
P_M=T\int_0^\infty\frac{dq}{4\pi}\ln(1+e^{-\beta(|q|-\mu_M)})+T\int_{-\infty}^\infty\frac{dq}{2\pi}\sum_{n>0}\ln(1+e^{-\beta(\sqrt{q^2+2neB}-\mu_M)})
\end{equation}
with $\mu_M=\mu+M\omega$.

\subsection{de Haas - van Alphen Oscillation}

As the standard derivation of the de Haas - van Alphen (dHvA) effect, the summation over the Landau level index $n$ can be carried out with the aid of the Poisson formula
\begin{equation}
\sum_{n=0}^\infty f(n)=\int_0^\infty f(n)dn+2{\rm Re}\sum_{l=1}^\infty\int_0^\infty f(n)e^{2i\pi ln}dx
\label{poisson}
\end{equation}
We have
\begin{equation}
F_M=F_{0M}+2{\rm Re}\sum_{l=1}^\infty F_{lM}
\label{decomp}
\end{equation}
where
\begin{equation}
F_{lM}=T\int_{-\infty}^\infty\frac{dq}{2\pi}\int_0^\infty dne^{i2\pi ln}\ln(1+e^{-\beta(\sqrt{q^2+2neB}-\mu_M)})
\end{equation}
The dHvA oscillation resides in the second term of (\ref{decomp}) and we shall focus on it.

Transforming the integration variables from $q, n$ to $q, \epsilon$ with $\epsilon=\sqrt{q^2+2neB}$, we find, via twice integration by part with respect to $\epsilon$, that
\begin{equation}
F_{lM}={\rm I}_{lM}+{\rm II}_{lM}+{\rm III}_{lM}
\end{equation}
for $l>0$, where
\begin{equation}
{\rm I}_{lM}=i\frac{eBT}{4\pi^2l}\int_{-\infty}^\infty dq\ln(1+e^{-\beta(q-\mu_M})),
\end{equation}
\begin{eqnarray}
	\begin{aligned}
{\rm II}_{lM}=\frac{eB}{4i\pi^2l}\sqrt{\frac{eB}{\pi l}}\int_{-\infty}^{\infty}dqe^{-i\frac{l\pi}{eB}q^2}\frac{\phi\left(\sqrt{\frac{l\pi}{eB}}|q|\right)}{e^{\beta(q-\mu_M)}+1}
\end{aligned}
\end{eqnarray}
and
\begin{eqnarray}
	\begin{aligned}
{\rm III}_{lM}=-\frac{eB}{4i\pi^2l}\sqrt{\frac{eB}{l\pi}}\int_0^\infty d\epsilon\phi\left(\sqrt{\frac{l\pi}{eB}}\epsilon\right)     \frac{\beta e^{\beta(\epsilon-\mu_M)}}{[e^{\beta(\epsilon-\mu_M)}+1]^2}\int_{-\epsilon}^\epsilon dqe^{-i\frac{l\pi}{eB}q^2}
\end{aligned} \end{eqnarray}
with
\begin{equation}
\phi(z)\equiv \int_z^\infty dxe^{ix^2}
\end{equation}
${\rm I}_{lM} $ is imaginary thereby does not contribute to (\ref{decomp}). Assuming the condition
\begin{equation}
T\ll\sqrt{eB}\ll\mu
\label{approx}
\end{equation}
the leading terms of ${\rm II}_{lM}$ and ${\rm III}_{lM}$ can be worked out and we ontain that
\begin{equation}
{\rm II}_{lM}=\frac{eB}{4\pi^3l^2}\left[\ln\left(\sqrt{\frac{4l\pi}{eB}}\mu_M\right)+\frac{1}{2}\gamma_E-i\frac{\pi}{4}\right]
\end{equation}
with $\gamma_E=0.5772...$ the Euler constant (See Appendix A for the derivation), and
\begin{equation}
{\rm III}_{lM}=-\frac{(eB)^{\frac{1}{2}}T}{4\pi}\frac{e^{i\left(\frac{l\pi^2}{eB}\mu_M^2-\frac{\pi}{4}\right)}}{l^{3/2}\sinh\frac{2l\pi^{2}T(\mu+M\omega)}{eB}}.
\end{equation}
where the integration formula
\begin{equation}
\int_{-\infty}^\infty dx\frac{e^{x+i\alpha}}{(e^x+1)^2}=\frac{\pi\alpha}{\sinh\pi\alpha}
\end{equation}
and the asymptotic form
\begin{equation}
\phi(z)=\frac{i}{2z}e^{iz^2}+...\hbox{    for $z\to\infty$}
\end{equation}
have been employed to reduce ${\rm III}_M$.
The dHvA osillation stems from ${\rm III}_M$. Summing over $M$, we end up with the dHvA term of the thermodynamic pressure under rotation, i.e.
\begin{equation}
  P_{\rm dHvA}\equiv\frac{1}{\pi R^2}\sum_{M>0}\left(2{\rm Re}\sum_{l=1}^\infty{\rm III}_{lM}\right)=-\frac{(eB)^{\frac{1}{2}}}{2\pi^{2}R^{2}}\sum_{l=1}^{\infty}\frac{1}{l^{3/2}}\sum_{M>0}\frac{\cos\left[\frac{l\pi}{eB}(\mu+M\omega)^{2}-\frac{\pi}{4}\right]}{\sinh\frac{2l\pi^{2}T(\mu+M\omega)}{eB}}
\label{dHvA_UR}
\end{equation}
In the absence of rotation, $\omega=0$, eq.(\ref{dHvA_UR}) becomes
\begin{equation}
P_{\rm dHvA}=-\frac{T(eB)^{\frac{3}{2}}}{4\pi^2}\sum_{l=1}^{\infty}\frac{1}{l^{3/2}}\frac{\cos\left[\frac{l\pi}{eB}\mu^2-\frac{\pi}{4}\right]}{\sinh\frac{2l\pi^{2}T\mu}{eB}}\to\frac{(eB)^{\frac{5}{2}}}{8\pi^4\mu}\sum_{l=1}^{\infty}\frac{1}{l^{5/2}}\cos\left[\frac{l\pi}{eB}\mu^2-\frac{\pi}{4}\right]
\label{w=0}
\end{equation}
in agreement with the expression derived from the Landau gauge.

Eq.(\ref{dHvA_UR}) can be further simplified at zero temperature, i.e.
\begin{equation}
P_{\rm dHvA}=-\frac{(eB)^{\frac{3}{2}}}{4\pi^{4}R^{2}}\sum_{M>0}\frac{1}{\mu+M\omega}\sum_{l=1}^{\infty}\frac{1}{l^{5/2}}\cos\left[\frac{l\pi}{eB}(\mu+M\omega)^{2}-\frac{\pi}{4}\right]\\
\end{equation}
The angular velocity and magnetic field considered throught this work satisfy the condition $\omega<<\sqrt{eB}$ and the summation over $M$ can be approximated by an integral. Consequently
\begin{eqnarray}\begin{aligned}
P_{\rm dHvA} \simeq & -\frac{(eB)^{\frac{3}{2}}}{4\pi^{4}R^{2}\omega}\int_{\mu}^{\mu+M_c\omega}dx\frac{1}{x}\sum_{l=1}^{\infty}\frac{1}{l^{5/2}}\cos\left(\frac{l\pi}{eB}x^{2}-\frac{\pi}{4}\right) \\
=& -\frac{(eB)^{\frac{3}{2}}}{8\sqrt{2}\pi^{4}R^{2}\omega}\sum_{l=1}^{\infty}\frac{1}{l^{5/2}}\left[\text{\ensuremath{\text{\ensuremath{\mathrm{Ci}}}\left(\frac{l\pi}{eB}(\mu+M_c\omega)^2\right)-\mathrm{Ci}}}\left(\frac{l\pi}{eB}\mu^2\right)\right. \\
&\left.+\mathrm{Si}\left(\frac{l\pi}{eB}(\mu+M_c\omega)^2 \right)-\mathrm{Si}\left(\frac{l\pi}{eB}\mu^2\right)\right] \\
\simeq & \frac{(eB)^{\frac{5}{2}}}{8\pi^{5}R^{2}\omega}\sum_{l=1}^{\infty}\frac{1}{l^{7/2}}\left[\frac{\sin\left(\frac{l\pi}{eB}\mu^2-\frac{\pi}{4}\right)}{\mu^{2}}-\frac{\sin\left(\frac{l\pi}{eB}(\mu+M_c\omega)^2-\frac{\pi}{4}\right)}{\left(\mu+M_c\omega\right)^{2}}\right]
\end{aligned}\label{dHvAUR}
\end{eqnarray}
where $\mathrm{Ci}(z)$ and $\mathrm{Si}(z)$ are cosine and sine integrals and the last step follows from their asymptotic forms for $z\gg1$, i.e.
\begin{equation}
\begin{cases}
\mathrm{Si}(z)\approx\frac{\pi}{2}-\frac{\cos z}{z}\\
\mathrm{Ci}(z)\approx\frac{\sin z}{z}
\end{cases}\label{sici}
\end{equation}
are employed in the last step. If the maximum rotation energy $M_c\omega$ dominates, i.e. $M_c\omega>>\mu$, the second term of (\ref{dHvAUR}) can be dropped and we have
\begin{equation}
P_{\rm dHvA} \simeq \frac{(eB)^{\frac{5}{2}}}{8\pi^{5}\mu^2R^{2}\omega}\sum_{l=1}^{\infty}\frac{1}{l^{7/2}}\sin\left(\frac{l\pi}{eB}\mu^2-\frac{\pi}{4}\right)
\label{neutronstar}
\end{equation}
and the uncertainty of $M_c$ does not contribute.

\subsection{Numerical Estimates}

As pointed out in the introduction, the rotation will lift the degeneracy of states within each Landau level and thereby reduce the de Haas - van Alphen oscillation. In this section, we shall estimate the amount of reduction using the parameters appropriate for two realistic rotating ultra-relativistic fermion system in a magnetic field, the quark matter core and a QGP droplet at high baryon density. Since the Fermi gas approximation of these two system tends to be poor and the condition of the latter syetem is highly transient, we are not attempting to model the two system. The signifinace of our result below is only in the sense of order of magnitude. For the ultra-relativistic system, we shall use $m_\pi=130$MeV as the scale of the chemical potential and temperature and $m_\pi^2=10^{14}$G as the scale of the magnetic field. The estimate of the impact of the de Haas - van Alphen effect in a non-relativistic fermion system is deferred to the next section.

\noindent{\it The quark matter core of a neutron star}
\begin{figure}[ht]
	\centering
	\includegraphics[width=3.12in,height=2.09in]{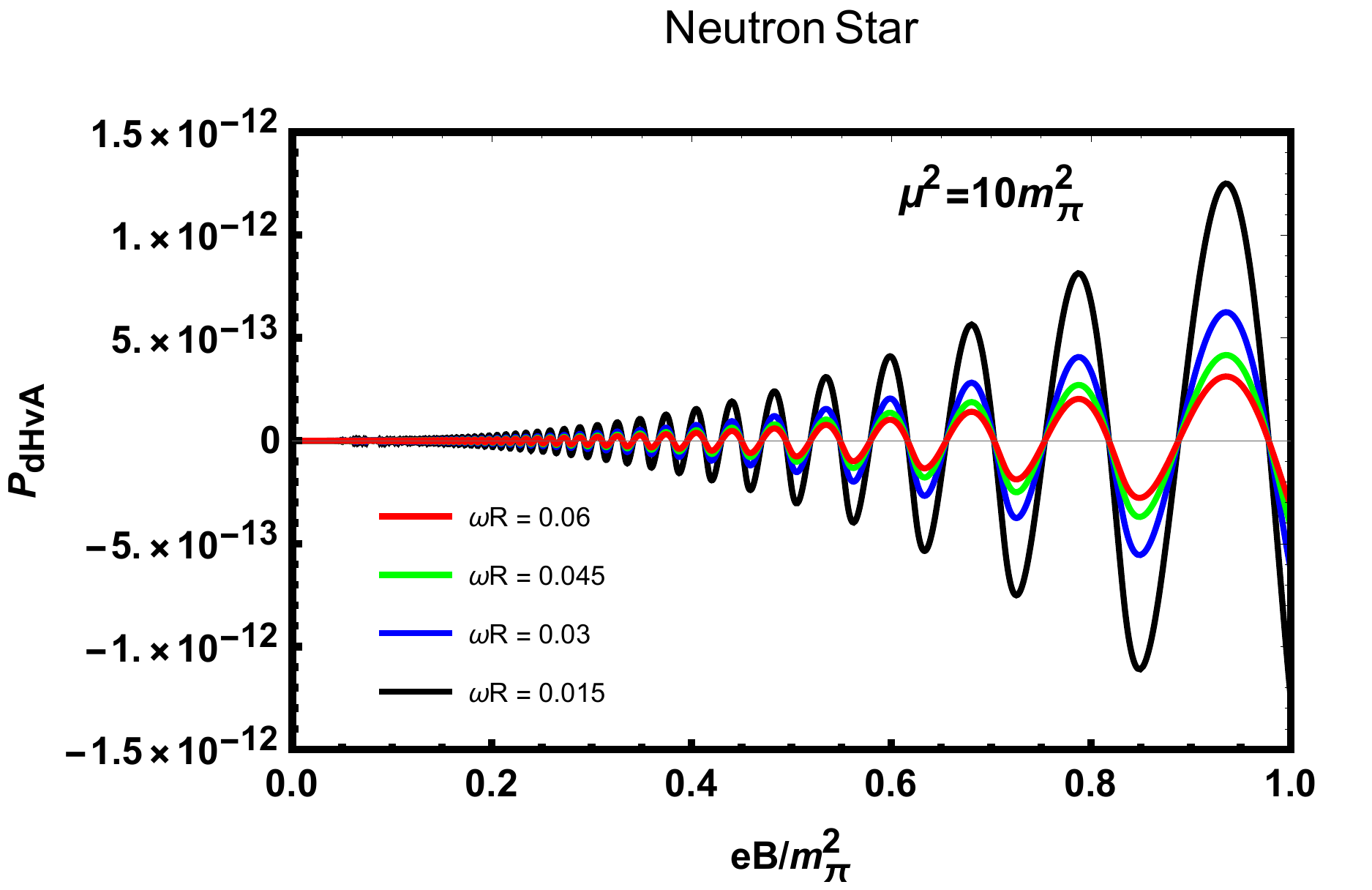}\\
	\caption{The oscillatory term of pressure $P_1$ as a function of magnetic field $\frac{eB}{m_\pi^2 }$. Here, $m_\pi=140$MeV, $R=1$km. }              \label{Neutron Star}
\end{figure}

The radius of a neutron star is of the order of 10km and we assume a quark matter core made of light flavors of smaller radius $R$ with a chemical potential of several hundreds of MeV, i.e. few times of pion's rest energy, $m_\pi$. The magnetic field inside a neutron star can reach as high as $10^{15}$G, i.e. $1.4\times 10^{-3}m_\pi^2$. For the fastest spinning neutron star, PSR J1748-2446ad, the frequency is 716Hz and the linear speed at the boundary of the core is $v\simeq 0.015$ (in the unit of the speed of light). Consequently
\begin{equation}
\frac{\mu}{M_c\omega}=\frac{\mu}{m_\pi}\cdot\frac{m_\pi^2}{eB}\cdot\frac{10^{-16}}{R(km)v}<<1
\end{equation}
\begin{equation}
\frac{P_{\rm dHvA}}{P_{\rm dHvA}\Vert_{\rm \omega=0}}\sim\frac{2}{\mu Rv}\simeq \frac{3.86\times10^{-16}}{\mu({\rm MeV})R({\rm km})v}
\end{equation}
for a typical neutron star. The approximation (\ref{neutronstar}) is valid and we estimate
 \begin{equation}
\frac{P_{\rm dHvA}}{P_{\rm dHvA}\Vert_{\rm \omega=0}}\sim\frac{2}{\mu Rv}\simeq \frac{3.86\times10^{-16}}{\mu({\rm MeV})R({\rm km})v}
\end{equation}
leading to huge suppression of dHvA oscillation.

\begin{figure}[ht]
	\centering
	\includegraphics[width=3.12in,height=2.09in]{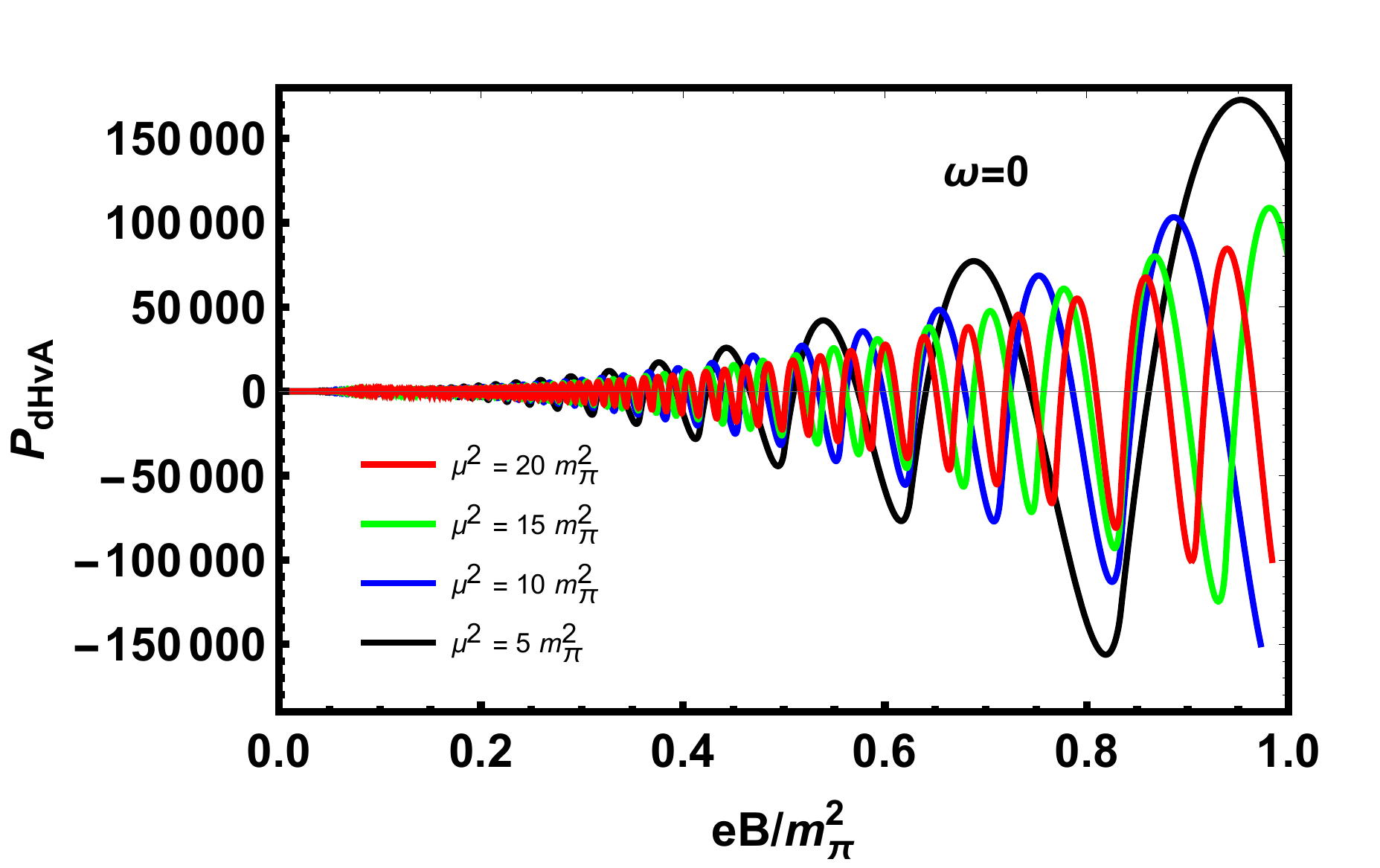}\\
	\caption{The oscillatory term of pressure $P_1$ as a function of magnetic field $\frac{eB}{m_\pi^2 }$. Here, $\omega=0$ and $T=0$.}              \label{0w}
\end{figure}

The thermodynamic pressure at $\mu^2=10m_\pi^2$ and zero temperature versus magnetic field $0<eB<0.01m_\pi^2$ is plotted in Fig.~{\ref{Neutron Star}} for several linear speeds at the boundary of the rotating quark matter core. As a benchmark, the thermodynamic pressure in the absence of rotation is displayed in Fig.~{\ref{0w}}. The parameters underlying both figures satisfy the approximation condition (\ref{approx}) for the analytic expressions. The effect is suppressed by 17 order of magnitude.

\noindent{\it A cold and dense QGP droplet}
\begin{figure}[ht]
	\centering
	\includegraphics[width=3.12in,height=2.09in]{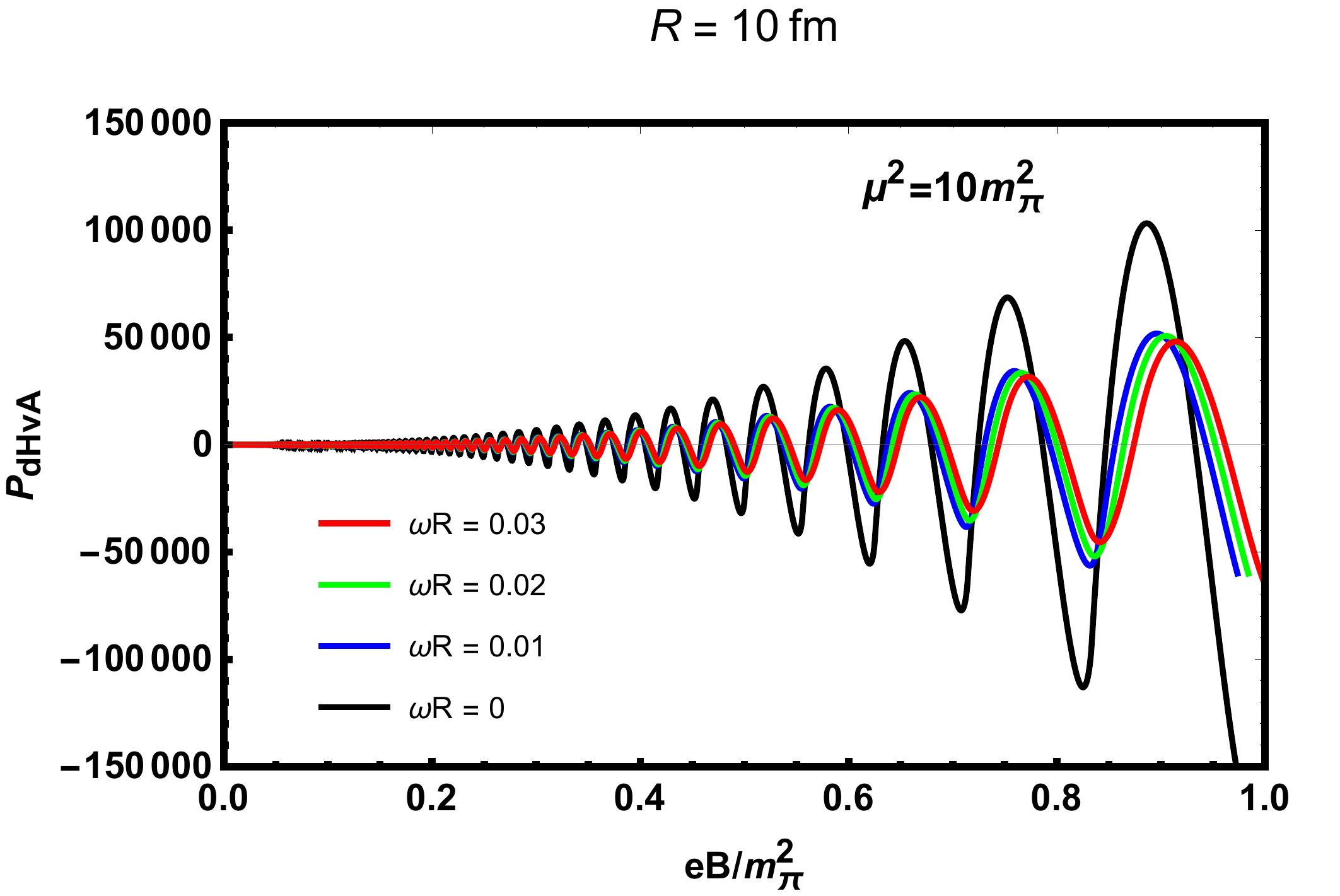}\\
	\caption{The oscillatory term of pressure $P_1$ as a function of magnetic field $\frac{eB}{m_\pi^2 }$. Here, we fix the chemical potential $\mu^2=10m_\pi^2$ and the radius is $R=10$fm.}              \label{10fm}
\end{figure}
\begin{figure}[ht]
	\centering
	\includegraphics[width=3.12in,height=2.09in]{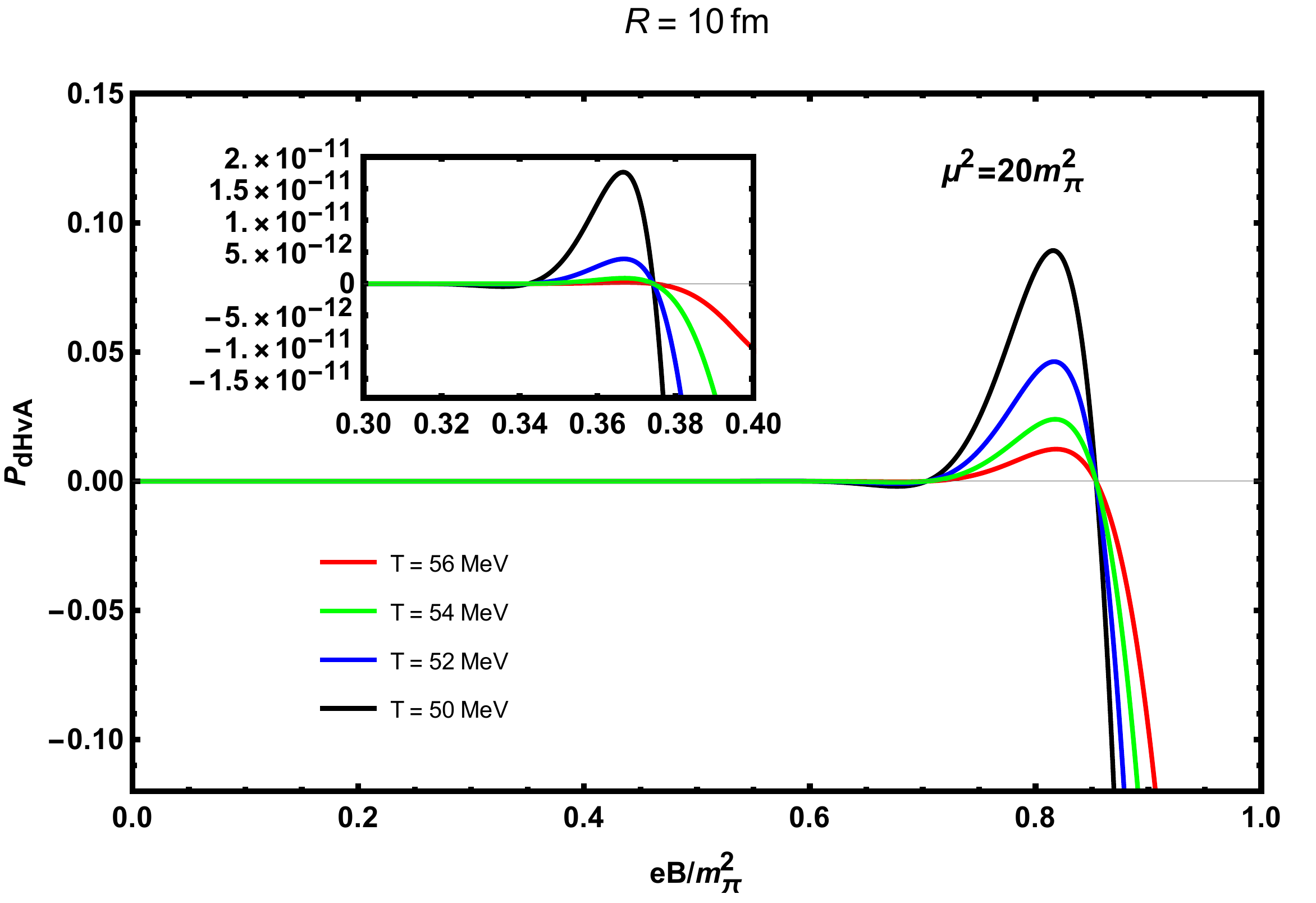}\\
	\caption{The oscillatory term of pressure $\frac{P_1}{(eB/m_\pi^2)^{30}}$ as a function of magnetic field $\frac{eB}{m_\pi^2 }$. Here, we fix the chemical potential $\mu^2=10m_\pi^2$, $v=0.01$ and the radius is $R=10$fm.}              \label{T}
\end{figure}
\begin{figure}[ht]
	\centering
	\includegraphics[width=3.12in,height=2.09in]{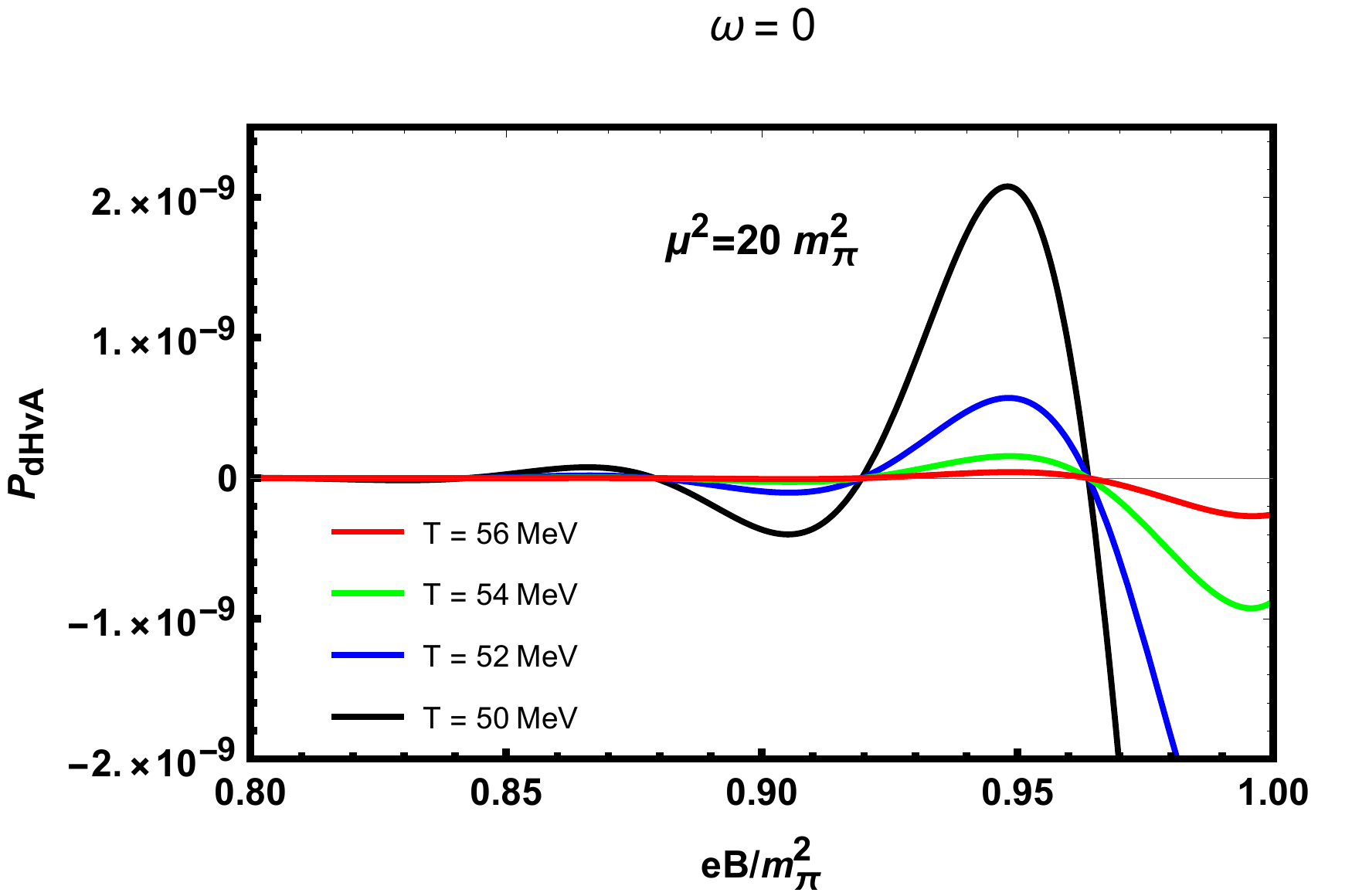}\\
	\caption{The oscillatory term of pressure $\frac{P_1}{(eB/m_\pi^2)^{30} }$ as a function of magnetic field $\frac{eB}{m_\pi^2 }$. Here, we fix the chemical potential $\mu^2=10m_\pi^2$, and $\omega=0$.}              \label{T-0w}
\end{figure}

The suppression of dHvA in a neutron star may be attributed to its large size. Let us switch to a cold and dense QGP droplet where the suppression of dHvA oscillation with the angular velocity becomes modest. The dHvA term of the thermodynamic pressure of eq.(\ref{dHvAUR}) for $R=10$fm versus the magnetic field at fixed chemical potential and temperature and is plotted for several angular velocity including $\omega=0$ in Fig.~{\ref{10fm}}. The same equation at fixed chemical potential and a nonzero angular velocity is plotted for several temperatures in Fig.~{\ref{T}}. The dHvA without rotation, eq.(\ref{w=0}) at the same chemical potential and the same set of tempertatures is plotted in Fig.~{\ref{T-0w}} for reference. Notice that the suppression of dHvA with temperature becomes milder with $\omega\ne 0$.  The selection of the size, chemical potential and the magnetic field is motivated by the conditions of the current heavy ion collisions in RHIC and LHC.

While the RHIC STAR fixed target experiment is expected to  generate QGP of lower energy and higher bayon density, i.e., closer to the density axis of the QCD phase diagram, there may still be a gap to meet the condition of the cold and dense QGP described above. Even it did, the rapid expansion would hinder the observability of the effect because of non-equilibrium. So our discussions here are highly speculative.

\section{Non Relativistic Fermi Gas}

The Hamiltonin of a non-relativistic electron reads
\begin{equation}
H=-\frac{1}{2m_e}({\vec\nabla}-ie{\vec A})^2+\frac{1}{2}\sigma_z\omega_B
\end{equation}
with the vector potential
\begin{equation}
\vec A=\frac{1}{2}B\hat z\times\vec r,
\end{equation}
where $\omega_B=eB/m_e$ is the cyclotron frequency and $\sigma_z={\rm diag.}(1,-1)$. The spectrum in cylindrical coordinates can be found in many textbook of quantum mechnics and are given by
\begin{equation}
E_{nmq\sigma}=\frac{q^2}{2m_e}+\left(n+\frac{m-|m|}{2}+\frac{1}{2}\right)\omega_B+\frac{1}{2}\sigma\omega_B
\label{NRLandau}
\end{equation}
where $q$ is the momentum along $z$-direction, $n=0,1,2,...$ are radial quantum number and m=0,$\pm 1$, $\pm 2$, ...,$\pm M_c$ are the z-component of the orbital angular momentum and $\sigma=\pm$ labels spin projections. The Landau levels correspond to $m\ge 0$ and are labeled by $n$. The corresponding wave function reads
\begin{equation}
\psi_{nmq\sigma}(\vec r)=\sqrt{\frac{n!eB}{2\pi(n+|m|)!L}}\zeta^{\frac{|m|}{2}}e^{-\frac{\zeta}{2}}L_n^{|m|}(\zeta)e^{i(m\phi+qz)}
\end{equation}
In a cylinder of finite radius, the thermodynamic approximation limits the azimuthal quantum number as (\ref{cutoff}), i.e.
\begin{equation}
|m|<m_c=[\frac{1}{2}eBR^2]>>1.
\label{NRcutoff}
\end{equation}
with an uncertainty $\delta m_c=O(1)$ as in the ultra-relativistic case.

\subsection{Thermodynamic Pressure and dHvA}

For a free non-relativistic electron gas, the dHvA can be extracted using the same Poisson formula (\ref{poisson}) as in most of the textbooks in solid state physics. Here we adapt a more elegant approach via Mellin transformation \cite{Peng}.

The thermodynamic pressure of the electron gas in a rotating cylindrical volume of radius $R$ and length $L_z$ reads
\begin{equation}
P=\frac{1}{\pi R^2}\sum_m P_m(\zeta_m)
\end{equation}
where
\begin{equation}
P_m(\zeta_m)=\frac{T}{L_z}\sum_{n,q,\sigma}\ln\left(1+\frac{1}{\zeta_m}e^{-\beta E_{qnm\sigma}}\right)
\end{equation}
with $\omega$ the angular velocity and
\begin{equation}
\zeta_m=e^{-\beta(\mu+m\omega)}
\end{equation}
The case of strong degeneracy corresponds to $\zeta_m<<1$. The Mellin transformation of the function $P_m(\zeta)$ with respect to $\zeta$ is given by
\begin{align}
  Q(s)=&\int_0^\infty d\zeta\zeta^{s-1}P_m(\zeta)\nonumber\\
      =& \frac{\pi T}{L_zs\sin\pi s}\sum_{n,q,\sigma}e^{-s\beta\left(E_{nmq\sigma}-\frac{1}{2}\sigma\omega\right)}
\end{align}
for $0<{\rm Re}s<1$. The last equality follows from an integration by part and the formula
\begin{equation}
\int_0^\infty dx\frac{x^{s-1}}{x+1}=\frac{\pi}{\sin\pi s}
\end{equation}
For the same reason as in the relativistic case, the contribution from $m<0$ is subleading in the thermodynamic approximation and we focus only on the branch $m\ge 0$ of the spectrum.
\begin{figure}[ht]
	\centering
	\includegraphics[width=3.12in,height=2.59in]{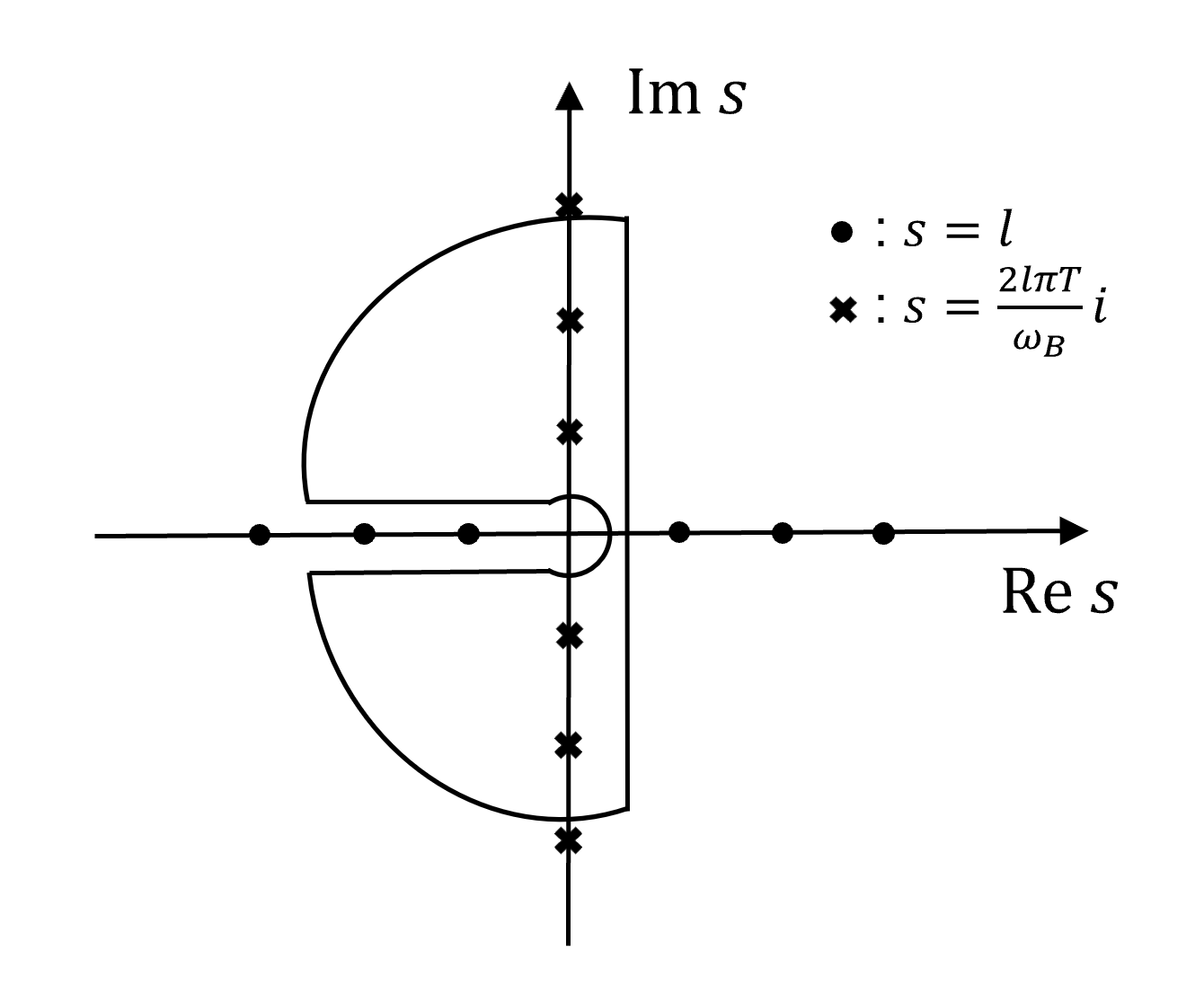}\\
	\caption{Contour integration~\cite{Peng}.}              \label{contour-integration}
\end{figure}
We have for $m\ge 0$
\begin{eqnarray}
Q(s) &=& \frac{\pi T}{L_zs\sin\pi s}\sum_q e^{-\frac{s\beta q^2}{2m_e}}\sum_{n,\sigma}e^{-\left(n+\frac{1}{2}\right)s\beta\omega_B-\frac{1}{2}\sigma s\beta(\omega_B-\omega)}\nonumber\\
&=& \frac{\pi T}{\lambda s^{3/2}\sin\pi s}\frac{\cosh \frac{1}{2}s\beta(\omega_B-\omega)}{\sinh\frac{1}{2}s\beta\omega_B}
\end{eqnarray}
where $\lambda=\sqrt{2\pi/(mT)}$ is the thermal wavelength. It follows from the Mellin inversion formula that
\begin{equation}
P_m(\zeta)=\int_{c-i\infty}^{c+i\infty}\frac{ds}{2\pi i}\zeta^{-s}Q(s)
\label{QS}
\end{equation}
with $0<c<1$. The integrand on the complex $s$-plane consists of a branch cut running along the negative real axis, poles along both real and imaginary axes, i.e.
\begin{equation}
s=l \qquad s=\frac{2l\pi T}{\omega_B}i
\end{equation}
with $l=0,\pm1,\pm2,...$. Closing the contour from the left as shown in Fig.\ref{contour-integration} for $\zeta<1$, we find
\begin{equation}
P_m(\zeta)={\rm I}_m(\zeta)+{\rm II}_m(\zeta)
\end{equation}
where ${\rm I}_m$ is the integral around the branch cut and ${\rm II}_m$ stems from the poles along the imaginary axis. The former contributes to the Landau diamagnetism and Pauli paramagnetism along with the Barnett effect and the latter gives rise to dHvA oscillation. Summing up the residues of the poles within the contour, we end up with
\begin{equation}
{\rm II}_m(\zeta_m)=\frac{2T}{\lambda}\sqrt{\frac{\omega_B}{2\pi T}}\sum_{l=1}^\infty\frac{1}{l^{3/2}}{\rm csch}\frac{2l\pi^2T}{\omega_B} \cos\frac{l\pi\omega}{\omega_B}\cos\left[\frac{2l\pi(\mu+m\omega)}{\omega_B}-\frac{\pi}{4}\right]
\end{equation}
Summing up the orbital angular momentum, we obtain that
\begin{eqnarray}
  P_{\rm dHvA} &=& \frac{1}{\pi R^2}\sum_{m=0}^{m_c}{\rm II}_m \nonumber\\
&=& -\frac{T(m_e\omega_B)^{1/2}}{\pi^2R^2}\sum_{l=1}^\infty\cos\frac{l\pi\omega}{\omega_B}\frac{\sin\left(\frac{2l\pi\mu}{\omega_B}-\frac{l\pi\omega}{\omega_B}-\frac{\pi}{4}\right)-\sin\left(\frac{2l\pi\mu}{\omega_B}+\frac{l\pi\omega}{\omega_B}-\frac{\pi}{4}+\frac{2l\pi m_c\omega}{\omega_B}\right)}{l^{3/2}\sinh\frac{2l\pi^2T}{\omega_B}\sin\frac{l\pi\omega}{\omega_B}}
\label{dHvA_NR}
\end{eqnarray}
Without rotation, $\omega=0$, the well-known dHvA formula
\begin{equation}
P_{\rm dHvA}|_{\omega=0}=-\frac{T(m_e\omega_B)^{3/2}}{2\pi^2}\sum_{l=1}^\infty\frac{1}{l^{3/2}}{\rm csch}\frac{2l\pi^2T}{\omega_B}\cos\left(\frac{2l\pi\mu}{\omega_B}-\frac{\pi}{4}\right)
\label{NRdHvA}
\end{equation}
emerges. At zero temperature, eq. (\ref{dHvA_NR}) becomes
\begin{eqnarray}
P_{\rm dHvA}|_{T=0} &=& -\frac{(m_e\omega_B)^{3/2}}{4\pi^4m_eR^2}\sum_{l=1}^\infty\cos\frac{l\pi\omega}{\omega_B}\frac{\sin\left(\frac{2l\pi\mu}{\omega_B}+\frac{l\pi\omega}{\omega_B}-\frac{\pi}{4}+l\pi m_e\omega R^2\right)-\sin\left(\frac{2l\pi\mu}{\omega_B}-\frac{l\pi\omega}{\omega_B}-\frac{\pi}{4}\right)}{l^{5/2}\sin\frac{l\pi\omega}{\omega_B}}\nonumber\\
&\simeq& -\frac{(m_e\omega_B)^{5/2}}{4\pi^5m_e^2\omega R^2}\sum_{l=1}^\infty\frac{1}{l^{7/2}}\left[\sin\left(\frac{2l\pi\mu}{\omega_B}-\frac{\pi}{4}+\frac{2l\pi m_c\omega}{\omega_B}\right)-\sin\left(\frac{2l\pi\mu}{\omega_B}-\frac{\pi}{4}\right)\right]
\label{NRdHvArot}
\end{eqnarray}
where the approximation $\omega<<\omega_B$ is made for the typical parameters in condensed matter physics. This expression is to be compared with the zero temperature limit of (\ref{NRdHvA}), i.e.
\begin{equation}
P_{\rm dHvA}|_{\omega=0}=-\frac{(m_e\omega_B)^{5/2}}{4\pi^4}\sum_{l=1}^\infty\frac{1}{l^{5/2}}\cos\left(\frac{2l\pi\mu}{\omega_B}-\frac{\pi}{4}\right).
\label{NRdHvA}
\end{equation}

At this point, it is interesting to compare the non-relativistic dHvA and the ultra-relativistic dHvA. As shown in eq.(\ref{NRLandau}), given $q$ and $\sigma$, the non-relativistic Landau levels (m>0) are equally spaced while the spacing between successive ultra-relativistic Landau levels in the upper equation of (\ref{landau}) decreases with the label $n$. Since the dHvA is sensitive to the energy levels around the chemical potential $\mu$, the amplitude of the oscillation is expected to be independent of $\mu$ in the non-relativistic case but decreases with $\mu$ in the ultra-relativistic case as reflected in the large $\mu$ suppression by $\sinh\frac{2l\pi^{2}T\mu}{eB}$ of (\ref{w=0}) in the latter case. When rotation is turned on, the effective chemical potential increases with the angular momentum quantum number. Consequently, the non-relativistic dHvA appears less vulnerable than the ultra-relativistic one.

\subsection{Numerical Estimates}

The electron gas in a good metal at room temperature, $T\sim 1/40$eV can be well approximated by a free Fermi in the strong degeneracy limit. The chemical potential is of $1\sim 10$eV, which makes $\mu/T\sim 40\sim 400>>1$ and the zero temperature approximation works well. For a magnetic field up to few Tesla's and an angular velocity is Hz, we have
\begin{equation}
\omega/\omega_B\simeq 5.57\times 10^{-12}\frac{\omega(Hz)}{B(Tesla)}
\end{equation}
justifying the approximation made in the (\ref{NRdHvArot}) for mechanical rotation achievable in laboratory.
The same condition also makes the contribution of the uncertainty in the angular momentum cutoff $m_c$ to the phase of the oscillation in (\ref{dHvA_NR}) and (\ref{NRdHvArot}) negligible.
The dHvA oscillation is expected to be significantly reduced when the largest rotation energy $m_c\omega$ within a Landau level exceeds the spacing between successive levels, $\omega_B$. With $R$ in cm, the linear velocity of the corcumference $v=\omega R$ in terms of cm/s, it follows from (\ref{NRcutoff}) that
\begin{equation}
\frac{m_c\omega}{\omega_B}\simeq 0.43Rv,
\label{critical}
\end{equation}
independent of the magnetic field.

\begin{figure}[ht]
	\centering
	\includegraphics[width=3.12in,height=2.09in]{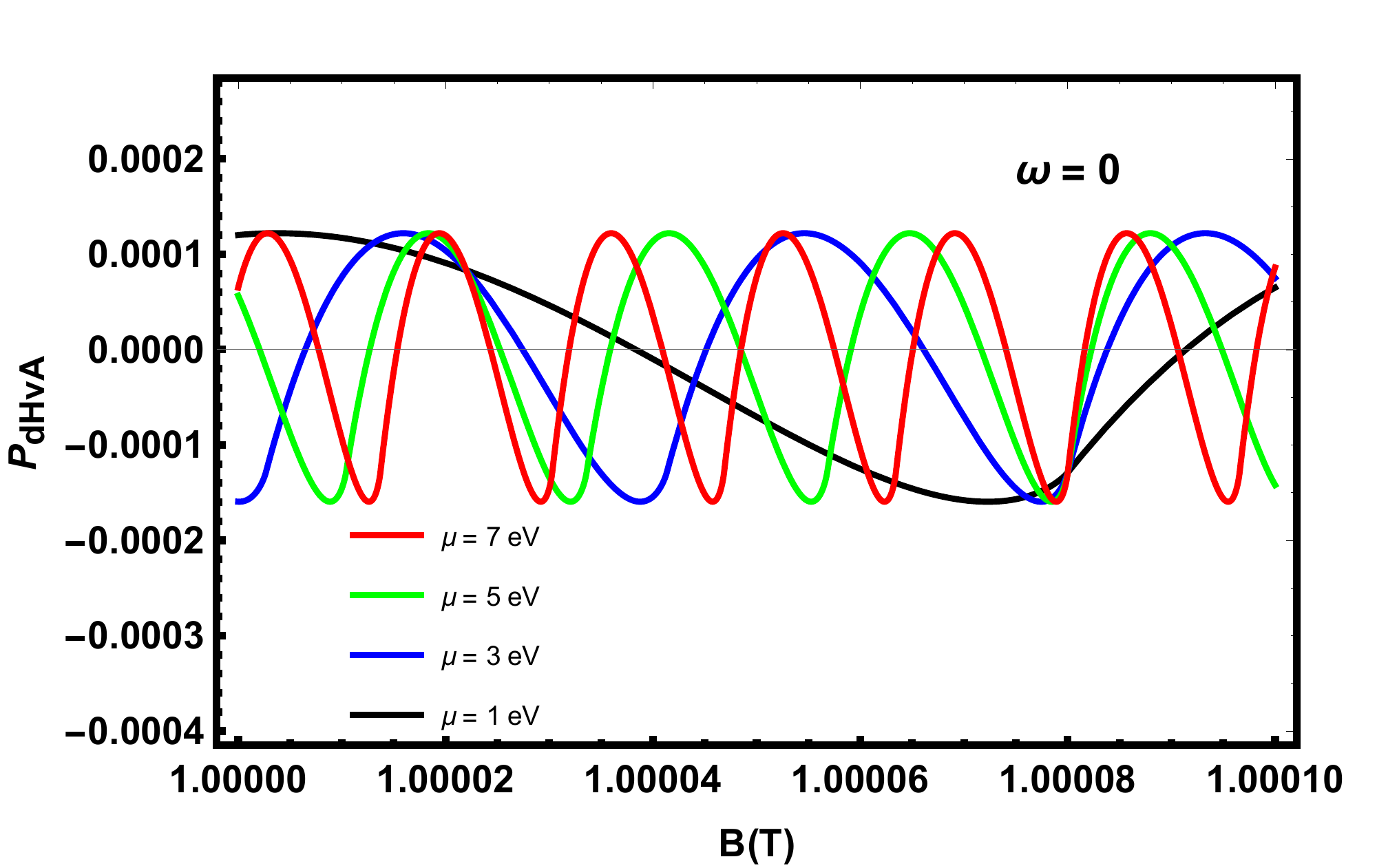}\\
	\caption{The oscillatory term of non-relativistic pressure $P_1$ as a function of magnetic field $B$ when $T=0$ and $\omega=0$.}              \label{NR-0T-0w}
\end{figure}

\begin{figure}[ht]
	\centering
	\includegraphics[width=3.12in,height=2.09in]{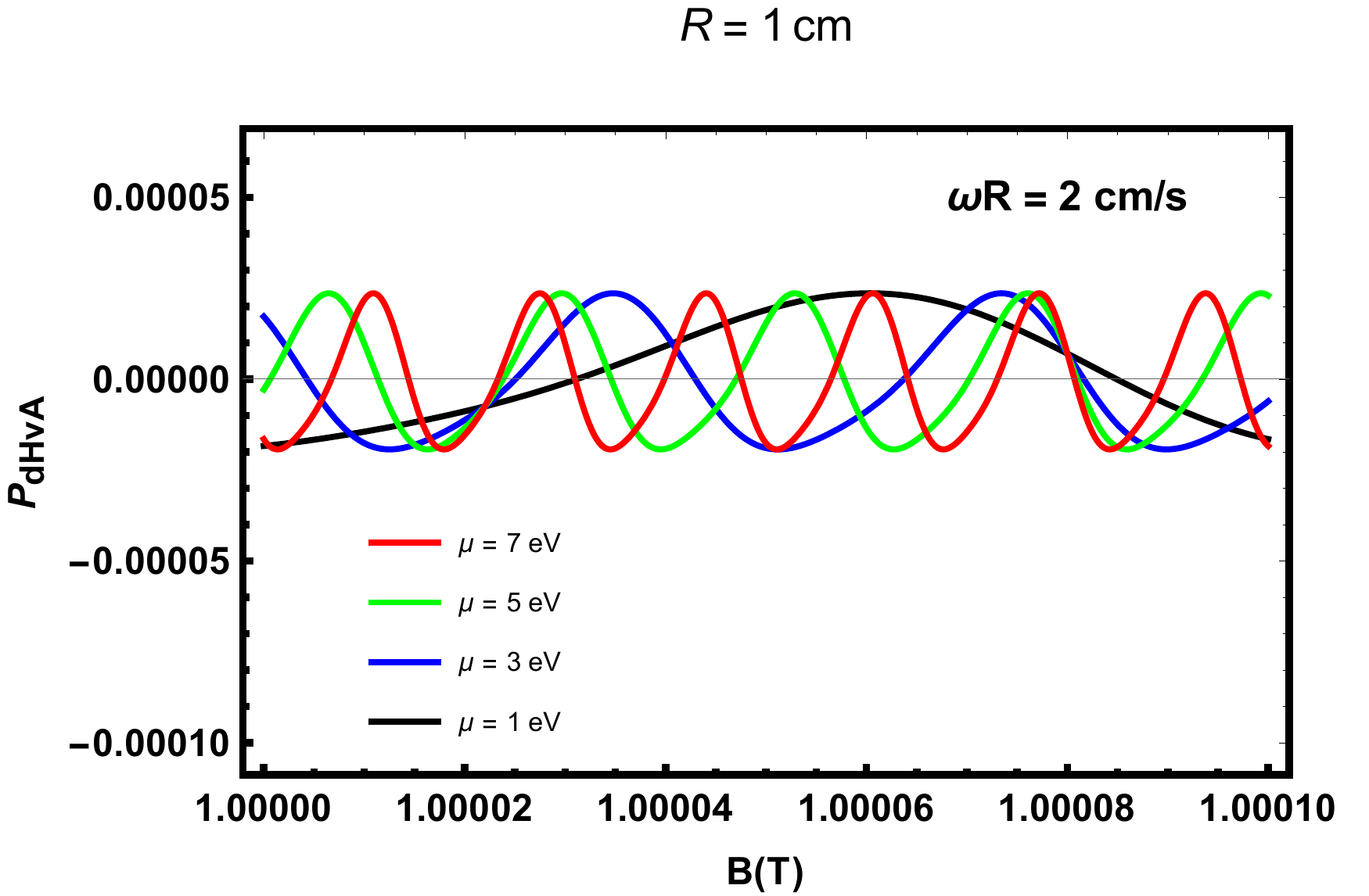}\\
	\caption{The oscillatory term of non-relativistic pressure $P_1$ as a function of magnetic field $B$ when $T=0$. Here, we fix $\omega R=2$cm/s and $R=1$ cm .}              \label{NR-0T-u}
\end{figure}

\begin{figure}[ht]
	\centering
	\includegraphics[width=3.12in,height=2.09in]{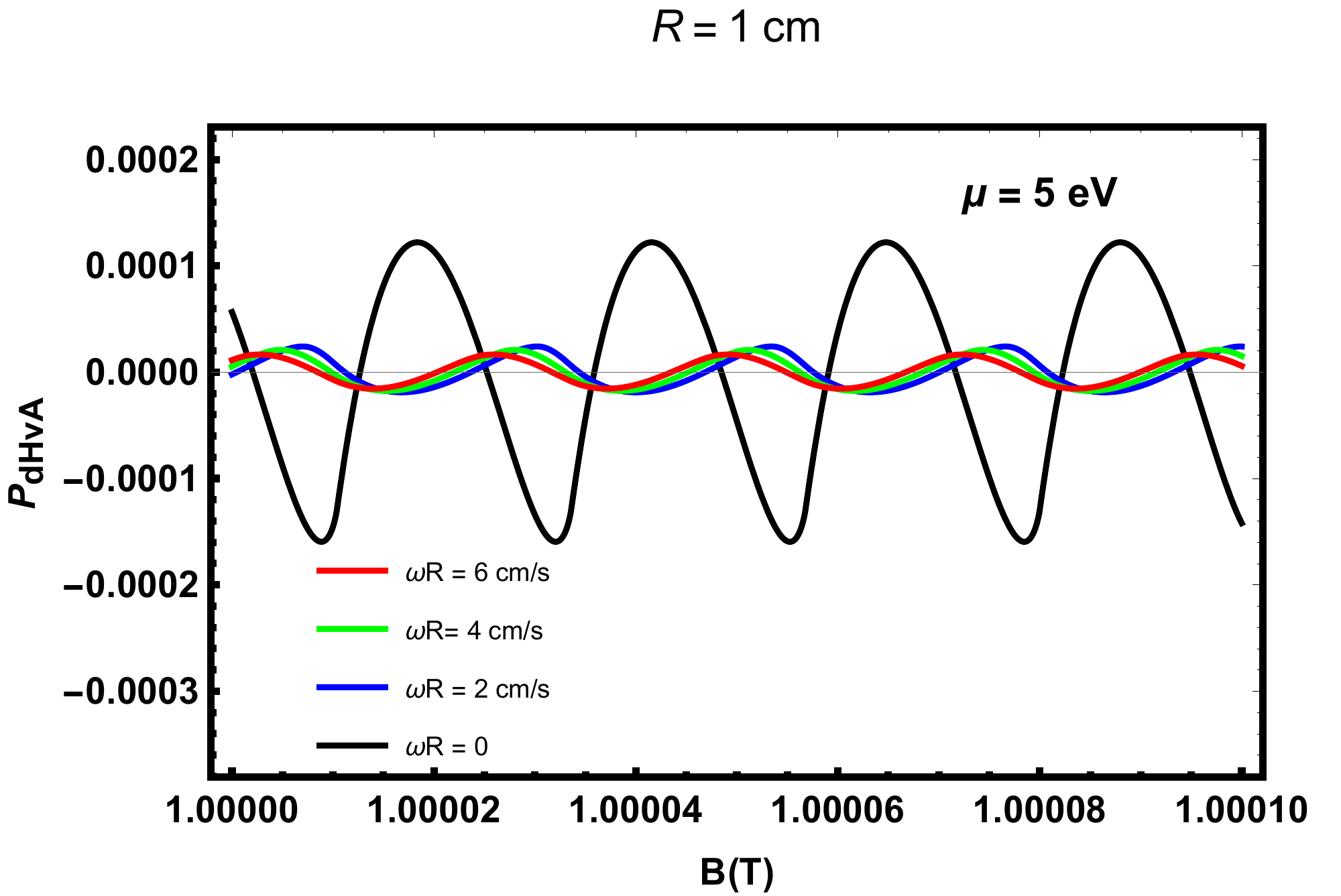}\\
	\caption{The oscillatory term of non-relativistic pressure $P_1$ as a function of magnetic field $B$ when $T=0$. Here, we fix the chemical potential $\mu=5$eV and the radius is $R=1$cm.}              \label{NR-1cm-0T}
\end{figure}

The dHvA term of the thermodynamic pressure of a strongly degenerate electron gas versus magnetic field for a long cylinder of radius $R=1$cm at $T=0$ is plotted in Fig.~{\ref{NR-0T-0w}}, Fig.~{\ref{NR-0T-u}} and Fig.~{\ref{NR-1cm-0T}}. The magnetic field varies in a small neighborhood of 1T and the angular velocity is taken such that RHS of (\ref{critical}) is of order one. The dHvA effect without rotation, eq.(\ref{NRdHvA}), for different chemical potentials is shown in Fig.~{\ref{NR-0T-0w}} sas benchmark. The parallel setup for $\omega R=2$cm/s, eq.(\ref{NRdHvArot}), is shown in Fig.~{\ref{NR-0T-u}} with similar profiles. More important is Fig.~{\ref{NR-1cm-0T}} where dHvA at different $\omega R$ is displayed and the suppression of the oscillation by rotation is evident.

\section{Concluding Remarks}

Let us recaptulate what we presented in preceding sections. We examined the robustness of the de Haas-van Alphen effect in a strongly degenerate Fermi gas under rotation. We derived the formula for dHvA oscillation in an long cylinder rotating about its axis in the ultra-relativistic limit and non-relativistic limit. As the macroscopic degeneracy of Landau levels is offset by rotation energy of states of different angular momentum within each Landau level. The amplitude of the scillation is reduced. The amount of reduction depends on the angular velocity $\omega$ and the radius of the cylinder $R$ and the oscillation is expected to become insignificant for sufficiently large $\omega$ and $R$. The ultra-relativistic dHvA appear more vulnerable than the non-relativistic one because of decreasing Landau level spacing with energy.

Applying the ultra-relativistic formula to estimate dHvA with typical parameters of a neutron star, and with typical parameters of a cold and dense QGP droplet, we noted that the dHvA oscillation is completely suppressed in the former case and remains in the latter. The non-relativistic formula, on the other hand showed that for a typical electron gas in a good metal, the variation of dHvA oscillation with angular velocity appears detectable, via magnetization and/or magnetic susceptibility.

As self-criticism, our approximation of the finite size effect by introducing the maximum angular momentum within a Landau level in (\ref{cutoff}) and (\ref{NRcutoff}) may be crude. Limited by the analytical tractability, the cylindrical shape of the system is not suitable to model a neutron star or a QGP droplet. Though the effect is expected to remain for a Fermi liquid, the strong correlation in quark matter may modify significantly the quantitative prediction. In this sense, our result is very preliminary.
\section*{Acknowledgments}
We thank Ren-Hong Fang for fruitful discussions. This work is supported by the National Key Research and Development Program of China (No. 2022YFA1604900). This work also  is supported by the National Natural Science Foundation of China (NSFC) under Grant Nos. 11735007, 11890711, 11890710, 12275104.
\appendix
\section{Appendix}
For $\mu>>T$, eq.(\ref{QS}) can be approximated as
\begin{equation}
{\rm II}_{lM}\simeq \frac{1}{2i\pi^2l}\sqrt{\frac{eB}{l\pi}}\int_0^{\mu_M}dqe^{-i\frac{l\pi}{eB}q^2}\phi\left(\sqrt{\frac{l\pi}{eB}}q\right)=\frac{eB}{2i\pi^3l^2}J
\label{II}
\end{equation}
where
\begin{equation}
J=\int_0^Kdxe^{-ix^2}\phi(x)=\int_0^Kdxe^{-ix^2}\int_x^\infty d\xi e^{i\xi^2}
\end{equation}
with $K=\sqrt{\frac{l\pi}{eB}}\mu_M$. Introducing $\xi=xt$, we find
\begin{eqnarray}
J &=& \int_0^Kdxe^{-ix^2}x\int_1^\infty dte^{it^2x^2}=\frac{1}{2i}\int_1^\infty dt\frac{e^{iK^2(t^2-1)}-1}{t^2-1} \nonumber\\
&=& -\frac{1}{2}K^2\int_1^\infty dte^{iK^2(t^2-1)}t\ln\frac{t-1}{t+1}
\end{eqnarray}
where the last equality follows from an integration by part. Introducing $z=t^2-1$, we have
\begin{equation}
J=-\frac{1}{4}K^2\int_0^\infty dze^{iK^2z}\ln\frac{\sqrt{z+1}-1}{\sqrt{z+1}+1}
\end{equation}
If follows from the Jordan lemma that integration path can be rotated to the imaginary axis on the $z-$ plane and we end up with
\begin{equation}
J=-\frac{i}{4}K^2\int_0^\infty dye^{-K^2y}\ln\frac{\sqrt{1+iy}-1}{\sqrt{1+iy}+1}
\end{equation}
For $K>>1$, we have
\begin{equation}
J\simeq-\frac{i}{4}K^2\int_0^\infty dye^{-K^2y}\ln\frac{iy}{4}
=\frac{i}{2}\left(\ln(2K)+\frac{1}{2}\gamma_E\right)+\frac{\pi}{8}
\end{equation}
This gives rise to RHS of (\ref{II}).

\bibliographystyle{unsrt}
\bibliography{de-Hass}

\end{document}